# AN OPTIMIZED ROUND ROBIN CPU SCHEDULING ALGORITHM WITH DYNAMIC TIME QUANTUM

Amar Ranjan Dash[1], Sandipta kumar Sahu[2] and Sanjay Kumar Samantra[3]

[1]Department of Computer Science, Berhampur University, Berhampur, India.
[2]Department of Computer Science, NIST, Berhampur, India.
[3]Department of Computer Science, NIST, Berhampur, India.

## ABSTRACT

*CPU scheduling is one of the most crucial operations performed by operating system. Different algorithms are available for CPU scheduling amongst them RR (Round Robin) is considered as optimal in time shared environment. The effectiveness of Round Robin completely depends on the choice of time quantum. In this paper a new CPU scheduling algorithm has been proposed, named as DABRR (Dynamic Average Burst Round Robin). That uses dynamic time quantum instead of static time quantum used in RR. The performance of the proposed algorithm is experimentally compared with traditional RR and some existing variants of RR. The results of our approach presented in this paper demonstrate improved performance in terms of average waiting time, average turnaround time, and context switching.*

## 1. INTRODUCTION

Operating systems are resource managers. The resources managed by Operating systems are hardware, storage units, input devices, output devices and data. Operating systems perform many functions such as implementing user interface, sharing hardware among users, facilitating input/output, accounting for resource usage, organizing data, etc. Process scheduling is one of the functions performed by Operating systems. CPU scheduling is the task of selecting a process from the ready queue and allocating the CPU to it. Whenever CPU becomes idle, a waiting process from ready queue is selected and CPU is allocated to that. The performance of the scheduling algorithm mainly depends on CPU utilization, throughput, turnaround time, waiting time, response time, and context switch.

Different CPU scheduling algorithms described by Abraham Silberschatz et al. [1], viz. FCFS (First Come First Served), SJF (Shortest Job First), Priority and RR (Round Robin). Neetu Goel et al. [2] make a comparative analysis of CPU scheduling algorithms with the concept of schedulers. Jayashree S. Somani et al. [3] also make a similar analysis but with their characteristics and applications. In FCFS, the process that requests the CPU first is allocated the CPU first. In SJF, the CPU is allocated to the process with smallest burst time. When the CPU becomes available, it is assigned to the process that has the smallest next CPU burst. If the next CPU bursts of two processes are the same, FCFS scheduling is used to break the tie. In priority scheduling algorithm a priority is associated with each process, and the CPU is allocated to the process with the highest priority. Equal priority processes are scheduled in FCFS order. A major problem with priority scheduling is starvation. In this scheduling some low priority processes wait indefinitely to get the CPU.

In RR a small unit of time is used which is called Time Quantum or Time slice. The CPU scheduler goes around the Ready Queue allocating the CPU to each process for a time interval up to 1 time quantum. If a process's CPU burst exceeds 1 time quantum, that process is pre-empted and is put back in the ready queue .If a new process arrives then it is added to the tail of the circular queue. Out of the above discussed algorithms RR provides better performance as compared to the others in case of a time sharing operating system. The performance of a scheduling algorithm depends upon the scheduling criteria viz. Turnaround time, Waiting time, Response time, CPU utilization, and throughput.

Turnaround time is the time interval from the submission time of a process to the completion time of a process. Waiting time is the sum of periods spent waiting in the ready queue. The time from the submission of a process until the first response is called Response time. The CPU utilization is the percentage of time CPU remains busy. The number of processes completed per unit time is called Throughput. Context switch is the process of swap-out the pre-executed process from CPU and swap-in a new process to CPU. Context switch is the number of times the process switches to get execute. A scheduling algorithm can be optimized by minimizing response time, waiting time and turnaround time and by maximizing CPU utilization, throughput.

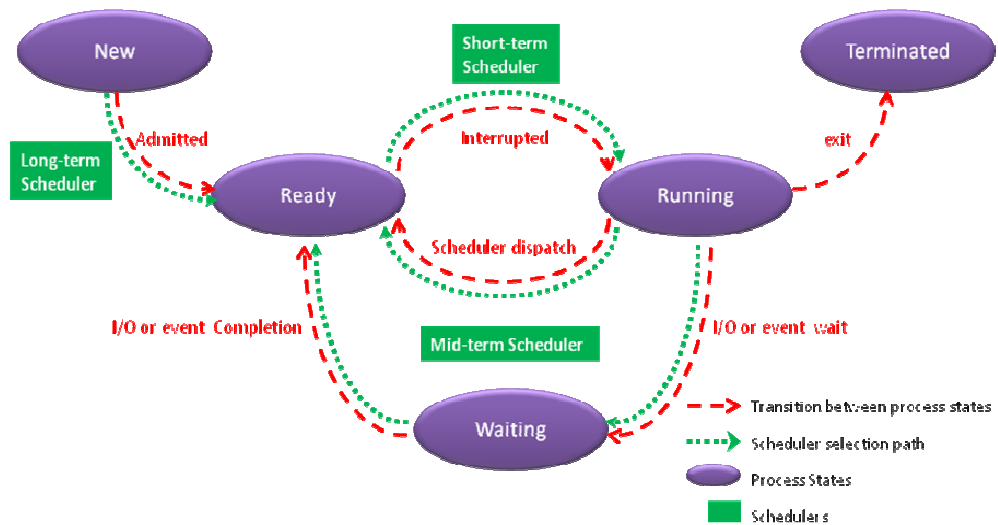

Figure 1: Different schedulers and Process states in CPU Scheduling.

The rest of the paper is organized as follows: In section 2, we describe the related works with a special emphasis on working procedure and dynamic time quantum selection procedure of different scheduling algorithms. Section 3 presents the proposed algorithm and its illustration. In section 4, we experimentally analyze the performance of seven scheduling algorithms, including our proposed algorithm, with six test cases. In section 5 we analyze the result obtained from our analysis. Section 6 provides the concluding remarks.

## 2. RELATED WORK

In recent time different approaches are used to increase the performance of CPU scheduling algorithm. Rakesh Kumar Yadav et al. [4] use the concept of SJF in RR algorithm. Ajit Singh et al. [5] combine the concept of SJF in RR algorithm. After each cycle they double the time quantum. Manish Kumar Mishra et al. [6] also merge the concept of Shortest Job First (SJF) with Round Robin (RR) to minimize the waiting time & turnaround time. After each complete cycle they chose the burst time of shortest process as new time quantum. Rishi Verma [7] calculate the time quantum after every cycle by subtracting the minimum burst time from maximum burst time. Neetu Goel et al. [8] take two dynamic numbers K (as time quantum) and F. During execution if the remaining burst time of process in execution is less than time-quantum/F, then the process continues its execution otherwise the process stops its execution and goes to the end of ready queue. M. Ramakrishna et al. [9] add the concept of priority scheduling in RR scheduling to optimize the Round Robin scheduling.

Rami J. Matarneh [10] proposes an algorithm SARR to improve the performance of Round Robin. In SARR for each cycle the median of burst time of the processes is calculated and used as time quantum. H.S.Behera et al. [11] also use similar type of algorithm. But they again rearrange the process during their execution. It select the process with lowest burst time, then process with highest burst time, then process with second lowest burst time, and so on.

# 3. OUR PROPOSAL

## 3.1. DABRR Algorithm

TQ:    Time Quantum
RQ:    Ready Queue
n:    number of process
Pi:    Process at $i^{th}$ index
i, j:    used as index of ready queue
TBT:    Total Burst Time
[1]    Arrange the processes in ascending order.
[2]    n = number of processes in RQ
[3]    i=0, TBT=0
[4]    Repeat step 5 and 6  till i < n
[5]      TBT += burst time of process Pi
[6]      i++
[7]    TQ = TBT/n
[8]    j = 0
[9]    Repeat from step 12 to 19 till j<n
[10]    if (burst time of Pi) <= TQ
[11]      Execute the process
[12]      Take the process out of RQ
[13]      n--
[14]    Else
[15]      Execute the process for a time interval up to 1 TQ
[16]      Burst time of Pi = Burst time of Pi – TQ
[17]      Add the process to ready queue for next round of execution
[18]    j++
[19]    If new process arrives
[20]    goto step 1
[21]    If RQ is not empty
[22]    goto step 2

## 3.2. Illustration:

In this section we analyzed the execution of the proposed algorithm. To demonstrate the proposed algorithm we have considered a ready queue with 5 processes p1, p2, p3, p4, p5. These processes are arrived at zero millisecond. The burst time of p1, p2, p3, p4, p5 are 15, 32, 102, 48, 29 milliseconds respectively. First the processes are arranged in ascending order of their burst time which provides the sequence p1, p5, p2, p4, p3. The time quantum is set equal to the mean of burst time of all 5 processes i.e. 45. After executing all processes for a time quantum of 45 millisecond execution of p1, p5, p2 get completed. So they are removed from the ready queue. After first cycle, the remaining burst time for p3 and p4 are 3 and 57 respectively. In the next cycle the new time quantum is set equal to the mean of the burst time of the processes in ready queue i.e. 30 and CPU is assigned to the processes for the new time quantum. After the second round the process p4 has finished execution and only process p5 remains in the ready queue with burst time 27 millisecond. As only one process is there in the ready queue so its burst time is directly chosen as time quantum and CPU is allocated to p5. According to our illustration the turnaround time for p1, p2, p3, p4, p5 are 15, 76, 226, 169, and 44 milliseconds respectively. The average turnaround time is 106. The waiting time for p1, p2, p3, p4, p5 are 0, 44, 124, 121, and 15 milliseconds respectively. The average waiting time is 60.8.

## 4. EXPERIMENTAL ANALYSIS

First we divide the problems into two types based on arrival time of processes (processes with zero arrival time and processes without zero arrival time). We further divide each into 3 more types based on the burst time of processes (in ascending order, descending order, & random order). We analyzed all the algorithms based on six situations. In each we have considered five processes with their arrival time and burst time. We have taken 25 as static time quantum for Round Robin.

Ajit Singh et al. [5] have proposed an algorithm in which after each cycle the time quantum is doubled. They haven't given any specific name for that algorithm. So during our experimental analysis we have considered it as R.P-5.As that paper has been referred at fifth place in our reference.

### 4.1. Assumptions

During analysis we have considered CPU bound processes only. In each test case 5 independent processes are analyzed in uni-processor environment. Corresponding burst time and arrival time of processes are known before execution. The context switch time of processes has been considered as zero. The time required for arranging the processes in ascending order also considered as zero.

### 4.2. With ZERO arrival time

#### 4.2.1. Ascending Order (Case I)

Table 1: Processes with Zero arrival time and Burst time in increasing order.

| Processes | Arrival time | Burst Time |
|-----------|--------------|------------|
| P1 | 0 | 40 |
| P2 | 0 | 55 |
| P3 | 0 | 60 |
| P4 | 0 | 90 |
| P5 | 0 | 102 |

**RR:**

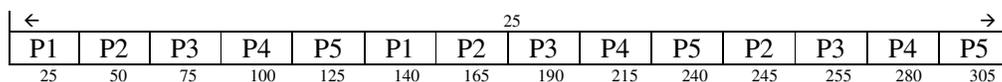

| | | | | | | | | | | | | | |
|---|---|---|---|---|---|---|---|---|---|---|---|---|---|
| P1 | P2 | P3 | P4 | P5 | P1 | P2 | P3 | P4 | P5 | P2 | P3 | P4 | P5 |
| 25 | 50 | 75 | 100 | 125 | 140 | 165 | 190 | 215 | 240 | 245 | 255 | 280 | 305 |

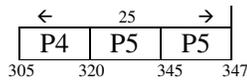

Average Turnaround time = (140+245+255+320+347)/5 = 1307/5 = 261.4
Average Waiting time
    = ((0+100) + (25+90+75) + (50+90+55) (75+90+40+25) + (100+90+40+15+0))/5
    = (100+190+195+230+245)/5 = 960/5 = 192

## DQRRR:

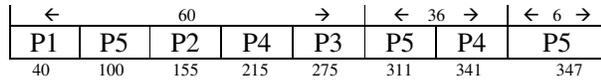

Average Turnaround time = (40+155+275+341+347)/5 = 1158/5 = 231.6
Average Waiting time
    = (0 + 100 + 215 + (155+96) + (40+175+30))/5
    = (0+100+215+251+245)/5 = 811/5 =162.2

## IRRVQ:

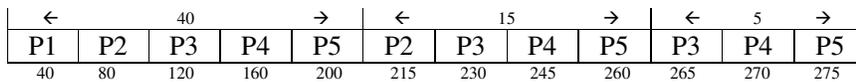

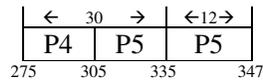

Average Turnaround time = (40+215+265+305+347)/5 = 1172/5 = 234.4
Average Waiting time = (0 + (40+120) + (80+95+30) + (120+70+20+5) + (160+45+10+30))/5
    = (0+160+205+215+245)/5 = 825/5 =165

## SARR:

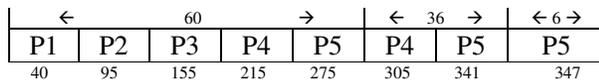

Average Turnaround time = (40+95+155+305+347)/5 = 942/5 = 188.4
Average Waiting time = (0 + 40 + 95 + (155+60) + (215+30+0))/5
    = (0+40+95+215+245)/5 = 595/5 = 119.0

## R.P-5:

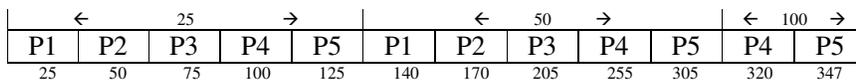

Average Turnaround time = (140+170+205+320+347)/5 = 1182/5 = 236.4
Average Waiting time = ((0+100) + (25+90) + (50+95) + (75+105+50) + (100+130+15))/5
    = (100+115+145+230+245)/5 = 835/5 =167

**MRR:**

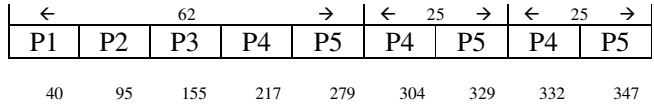

Average Turnaround time = (40+95+155+332+347)/5 = 969/5 = 193.8
Average Waiting time = (0 + 40 + 95 + (155+62+25) + (217+25+3))/5
    = (0+40+95+242+245)/5 = 622/5 =124.4

**DABRR:**

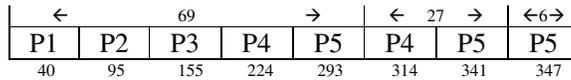

Average Turnaround time = (40+95+155+314+347)/5 = 951/5 = 190.2
Average Waiting time = (0 + 40 + 95 + (155+69) + (224+21))/5
    = (0+40+95+224+245)/5 = 604/5 = 120.8

Table 2: Comparison between RR, DQRRR, IRRVQ, SARR, RP-5, MRR, DABRR for case-I.

| Algorithm | Round Robin | DQRRR | IRRVQ | SARR | RP-5 | MRR | DABRR |
|---|---|---|---|---|---|---|---|
| Time Quantum | 25 | 60,36,6 | 40,15,5, 30,12 | 60, 36, 6 | 25,50, 100 | 62, 25, 25 | 69,27,6 |
| Context Switch | 16 | 7 | 14 | 7 | 11 | 8 | 7 |
| Average Waiting time | 192 | 162.2 | 165 | 119 | 167 | 124.4 | 120.8 |
| Average Turn Around Time | 261.4 | 231.6 | 234.4 | 188.4 | 236.4 | 193.8 | 190.2 |

### 4.2.2. Descending Order (Case II)

Table 3:  Processes with zero arrival time and Burst time in decreasing order.

| Processes | Arrival time | Burst Time |
|---|---|---|
| P1 | 0 | 105 |
| P2 | 0 | 85 |
| P3 | 0 | 55 |
| P4 | 0 | 43 |
| P5 | 0 | 35 |

**RR:**

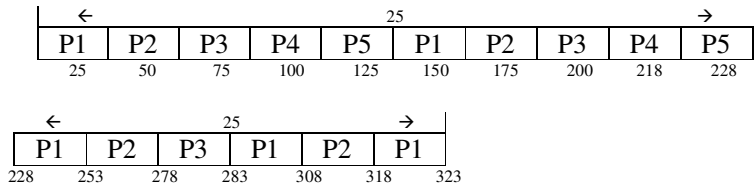

Average Turnaround time = (323+318+283+218+228)/5 = 1370/5 = 274

Average Waiting time
= ((0+100+78+30+10) + (25+100+78+30) + (50+100+78) + (75+100) + (100+93))/5
= (218+233+228+175+193)/5 = 1047/5 = 209.4

## DQRRR:

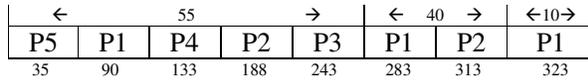

Average Turnaround time = (323+313+243+133+35)/5 = 1047/5 = 209.4

Average Waiting time = ((35+153+30) + (133+95) +188+90+0)/5
= (218+228+188+90+0)/5 = 724/5 =144.8

## IRRVQ:

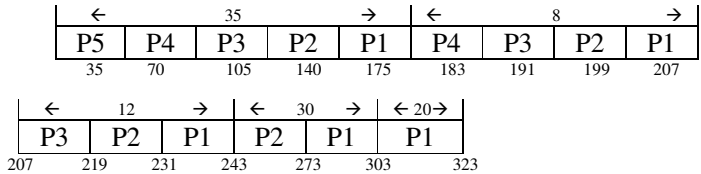

Average Turnaround time = (323+273+219+183+35)/5 = 1033/5 = 206.6
Average Waiting time
= ((140+24+24+30+0) + (105+51+20+12) + (70+78+16) + (35+105) +0)/5
= (218+188+164+140+0)/5 = 710/5 =142

## SARR:

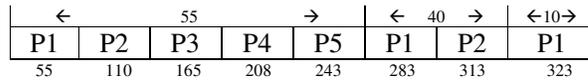

Average Turnaround time = (323+313+165+208+243)/5 = 1252/5 = 250.4
Average Waiting time = ((0+188+30) + (55+173) +110+165+208)/5
= (218+228+110+165+208)/5 = 929/5 =185.8

## R.P-5:

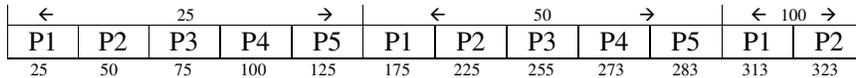

Average Turnaround time = (313+323+255+273+283)/5 = 1447/5 = 289.4
Average Waiting time = ((0+100+108) + (25+125+88) + (50+150) + (75+155) + (100+148))/5
= (208+238+200+230+248)/5 = 1124/5 =224.8

## MRR:

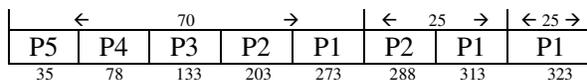

Average Turnaround time = (323+288+133+78+35)/5 = 857/5 = 171.4

Average Waiting time = ((203+15+0) + (133+70) + 78 + 35 + 0))/5
    = (218+203+78+35+0)/5 = 534/5 = 106.8

**DABRR:**

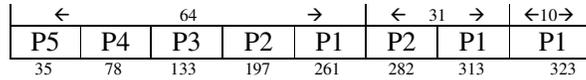

| P5 | P4 | P3 | P2 | P1 | P2 | P1 | P1 |
|----|----|----|----|----|----|----|----|
| 35 | 78 | 133 | 197 | 261 | 282 | 313 | 323 |

Average Turnaround time = (323+282+133+78+35)/5 = 851/5 = 170.2
Average Waiting time = ( (197+21+0) + (133+64)+78+35+0)/5
        = (218+197+78+35+0)/5 = 528/5 = 105.6

Table 4: Comparison between RR, DQRRR, IRRVQ, SARR, RP-5, MRR, DABRR for case-II.

| Algorithm | Round Robin | DQRRR | IRRVQ | SARR | RP-5 | MRR | DABRR |
|-----------|-------------|-------|-------|------|------|-----|-------|
| Time Quantum | 25 | 55,40,10 | 35,8,12, 30,20 | 55, 40, 10 | 25,50, 100 | 70, 25, 25 | 64,31,10 |
| Context Switch | 15 | 7 | 14 | 7 | 11 | 7 | 7 |
| Average Waiting time | 209.4 | 144.8 | 142 | 185.8 | 224.8 | 106.8 | 105.6 |
| Average Turn Around Time | 274 | 209.4 | 206.6 | 250.4 | 289.4 | 171.4 | 170.2 |

### 4.2.3. Random Order (Case III)

Table 5:  Processes with zero arrival time and Burst time in random order.

| Processes | Arrival time | Burst Time |
|-----------|--------------|------------|
| P1 | 0 | 105 |
| P2 | 0 | 60 |
| P3 | 0 | 120 |
| P4 | 0 | 48 |
| P5 | 0 | 75 |

**RR:**

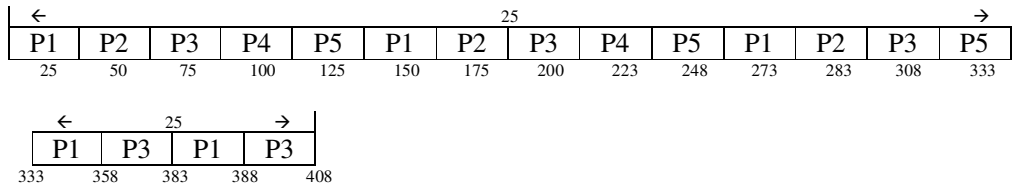

| P1 | P2 | P3 | P4 | P5 | P1 | P2 | P3 | P4 | P5 | P1 | P2 | P3 | P5 |
|----|----|----|----|----|----|----|----|----|----|----|----|----|----|
| 25 | 50 | 75 | 100 | 125 | 150 | 175 | 200 | 223 | 248 | 273 | 283 | 308 | 333 |

| P1 | P3 | P1 | P3 |
|----|----|----|----|
| 333 | 358 | 383 | 388 | 408 |

Average Turnaround time = (388+283+408+223+333)/5 = 1635/5 = 327
Average Waiting time
= ((0+100+98+60+25) + (25+100+98) + (50+100+83+50+5) + (75+100) + (100+98+60))/5
= (283+223+288+175+258)/5 = 1227/5 = 245.4

## DQRRR:

| | | ← | 75 | → | | ← | 37 | → | ←8→ |
|---|---|---|---|---|---|---|---|---|---|
| P4 | P3 | P2 | P1 | P5 | P3 | P1 | P3 |
| 48 | 123 | 183 | 258 | 333 | 370 | 400 | 408 |

Average Turnaround time = (400+183+408+48+333)/5 = 1372/5 = 274.4
Average Waiting time = ((183+112) + 123+ (48+210+30) + 0 + 258)/5
  = (295+123+288+0+258)/5 = 964/5 =192.8

## IRRVQ:

| | ← | 48 | → | ← | 12 | → | ← | 15 | → |
|---|---|---|---|---|---|---|---|---|---|
| P4 | P2 | P5 | P1 | P3 | P2 | P5 | P1 | P3 | P5 | P1 | P3 |
| 48 | 96 | 144 | 192 | 240 | 252 | 264 | 276 | 288 | 303 | 318 | 333 |

| ← | 30 | → | ←15→ |
|---|---|---|---|
| P1 | P3 | P3 |
| 333 | 363 | 393 | 408 |

Average Turnaround time = (363+252+408+48+303)/5 = 1374/5 = 274.8
Average Waiting time
  = ((144+72+27+15) + (48+144) + (192+36+30+30+0) + 0 + (96+108+24))/5
  = (258+192+288+0+228)/5 = 966/5 =193.2

## SARR:

| ← | 120 | → |
|---|---|---|
| P1 | P2 | P3 | P4 | P5 |
| 105 | 165 | 285 | 333 | 408 |

Average Turnaround time = (105+165+285+333+408)/5 = 1296/5 = 259.2
Average Waiting time = (0+105+165+285+333)/5 = 888/5 = 177.6

## R.P-5:

| | ← | 25 | → | ← | 50 | → | ← | 100 | → |
|---|---|---|---|---|---|---|---|---|---|
| P1 | P2 | P3 | P4 | P5 | P1 | P2 | P3 | P4 | P5 | P1 | P3 |
| 25 | 50 | 75 | 100 | 125 | 175 | 210 | 260 | 283 | 333 | 363 | 408 |

Average Turnaround time = (363+210+408+283+333)/5 = 1597/5 = 319.4
Average Waiting time
  = ((0+100+158) + (25+125) + (50+135+103) + (75+160) + (100+158))/5
  = (258+150+288+235+258)/5 = 1189/5 = 237.8

## MRR:

| | ← | 72 | → | ← | 45 | → | ←25→ |
|---|---|---|---|---|---|---|---|
| P4 | P2 | P5 | P1 | P3 | P5 | P1 | P3 | P3 |
| 48 | 108 | 180 | 252 | 324 | 327 | 360 | 405 | 408 |

Average Turnaround time = (360+108+408+48+327)/5 = 1251/5 = 250.2
Average Waiting time = ((180+75) + 48 + (252+36+0) + 0 + (108+144))/5
  = (255+48+288+0+252)/5 = 843/5 = 168.6

**DABRR:**

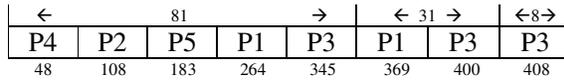

Average Turnaround time = (369+108+408+48+183)/5 = 1116/5 = 223.2
Average Waiting time = ((183+81) + 48 + (264+24+0) + 0 + 108)/5
      = (264+48+288+0+108)/5 = 708/5 = 141.6

Table 6: Comparison between RR, DQRRR, IRRVQ, SARR, RP-5, MRR, DABRR for case-III.

| Algorithm | Round Robin | DQRRR | IRRVQ | SARR | RP-5 | MRR | DABRR |
|---|---|---|---|---|---|---|---|
| Time Quantum | 25 | 75,37,8 | 48,12,15, 30,15 | 120 | 25,50, 100 | 72, 45, 25 | 81,31,8 |
| Context Switch | 17 | 7 | 14 | 4 | 11 | 8 | 7 |
| Average Waiting time | 245.4 | 192.8 | 193.2 | 177.6 | 237.8 | 168.6 | 141.6 |
| Average Turn Around Time | 327 | 274.4 | 274.8 | 259.2 | 319.4 | 250.2 | 223.2 |

### 4.2.4. Comparison

This section provides the comparative analysis of seven algorithms on the basis of their resulted waiting time and turnaround time. Table 7 shows the performance analysis of seven algorithms by summarizing the waiting time and turnaround time resulted from case I, II, III.

Table 7: Comparison of six algorithms by aggregating their performance measure from case I, II, III.

| | \multicolumn{4}{c}{Context Switch} | | | | Waiting Time | | | | Turnaround Time | | | |
|---|---|---|---|---|---|---|---|---|---|---|---|---|---|
| | Ascending Burst Time | Descending Burst Time | Random Burst Time | TOTAL | Ascending Burst Time | Descending Burst Time | Random Burst Time | TOTAL | Ascending Burst Time | Descending Burst Time | Random Burst Time | TOTAL |
| **RR** | 16 | 15 | 17 | **48** | 192.00 | 209.40 | 245.40 | **646.80** | 261.40 | 274.00 | 327.00 | **862.40** |
| **DQRRR** | 7 | 7 | 7 | **21** | 162.20 | 144.80 | 192.80 | **499.80** | 231.60 | 209.40 | 274.40 | **715.40** |
| **IRRVQ** | 14 | 14 | 14 | **42** | 165.00 | 142.00 | 193.20 | **500.20** | 234.40 | 206.60 | 274.80 | **715.80** |
| **SARR** | 7 | 7 | 4 | **18** | 119.00 | 185.80 | 177.60 | **482.40** | 188.40 | 250.40 | 259.20 | **698.00** |
| **RP-5** | 11 | 11 | 11 | **33** | 167.00 | 224.80 | 237.80 | **629.60** | 236.40 | 289.40 | 319.40 | **845.20** |
| **MRR** | 8 | 7 | 8 | **23** | 124.40 | 106.80 | 168.60 | **399.80** | 193.80 | 171.40 | 250.20 | **615.40** |
| **DABRR** | 7 | 7 | 7 | **21** | 120.80 | 105.60 | 141.60 | **368.00** | 190.20 | 170.20 | 223.20 | **583.60** |

The table title row: **With 0 Arival Time**

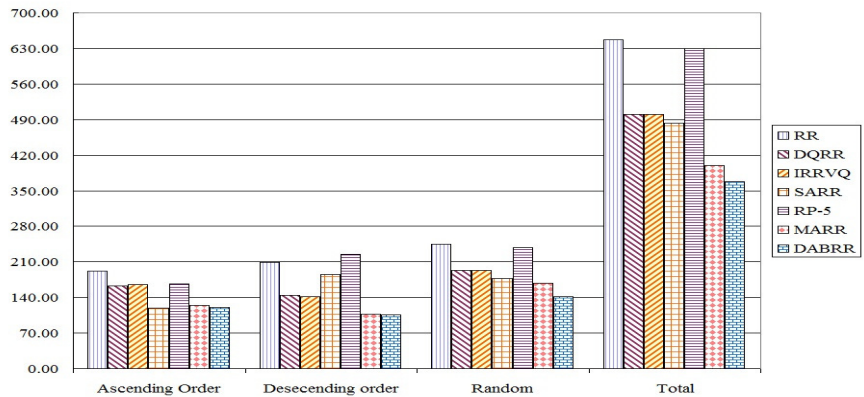

Figure 2: Waiting time for processes with 0 arrival time.

Figure 2 & 3 show the graphical analysis of the waiting time and turnaround time respectively of all algorithms based on case I, II, III. By evaluating these two graphs we conclude that our algorithm performs better than other six algorithms in case of processes arrived at zero arrival time.

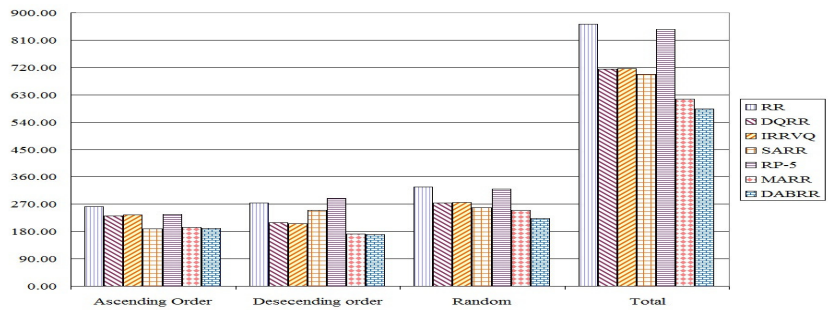

Figure 3: Turnaround time for processes with 0 arrival time.

## 4.3. Without ZERO arrival time

### 4.3.1. Ascending Order (Case IV)

Table 8: Processes without Zero arrival time and Burst time in increasing order.

| Processes | Arrival time | Burst Time |
|-----------|--------------|------------|
| P1 | 0 | 27 |
| P2 | 3 | 32 |
| P3 | 5 | 55 |
| P4 | 7 | 82 |
| P5 | 9 | 110 |

**RR:**

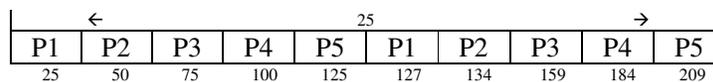

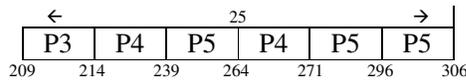

Average Turnaround time = (127+ (134-3) + (214-5) + (271-7) + (306-9))/5 = 1028/5 = 205.6
Average Waiting time
= ((0+100) + (25+77-3) + (50+59+50-5) + (75+59+30+25-7) + (100+59+30+7+0-9))/5
= (100+99+154+182+187)/5 = 722/5 = 144.4

## DQRRR:

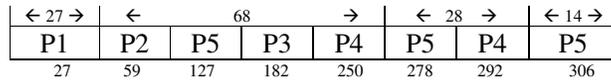

Average Turnaround time = (27+ (59-3) + (182-5) + (292-7) + (306-9))/5 = 842/5 = 168.4
Average Waiting time
    = (0 + (27-3) + (127-5) + (182+28-7) + (59+123+14-9))/5
    = (0+24+122+203+187)/5 = 536/5 =107.2

## IRRVQ:

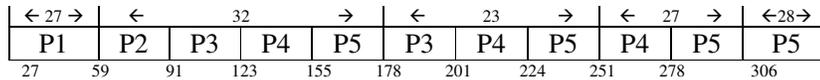

Average Turnaround time = (27+ (59-3) + (178-5) + (251-7) + (306-9))/5 = 797/5 = 159.4
Average Waiting time
    = (0 + (27-3) + (59+64-5) + (91+55+23-7) + (123+46+27+0-9))/5
    = (0+24+118+162+187)/5 = 491/5 = 98.2

## SARR:

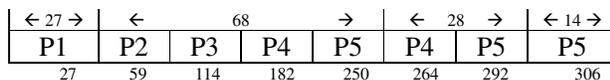

Average Turnaround time = (27+ (59-3) + (114-5) + (264-7) + (306-9))/5 = 746/5 = 149.2
Average Waiting time
    = (0 + (27-3) + (59-5) + (114+68-7) + (182+14+0-9))/5
    = (0+24+54+175+187)/5 = 440/5 = 88

## R.P-5:

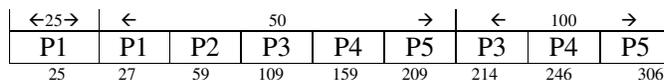

Average Turnaround time = (27+ (59-3) + (214-5) + (246-7) + (306-9))/5 = 828/5 = 165.6
Average Waiting time
    = (0 + (27-3) + (59+100-5) + (109+55-7) + (159+37-9))/5
    = (0+24+154+157+187)/5 = 522/5 =104.4

**MRR:**

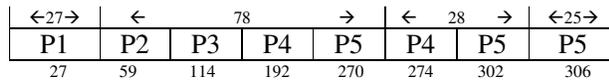

Average Turnaround time = (27+ (59-3) + (114-5) + (274-7) + (306-9))/5 = 756/5 = 151.2
Average Waiting time
$\quad$ = (0 + (27-3) + (59-5) + (114+78-7) + (192+4+0-9))/5
$\quad$ = (0+24+54+185+187)/5 = 450/5 = 90

**DABRR:**

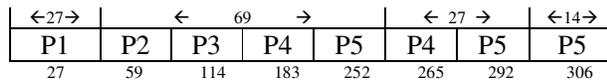

Average Turnaround time = (27+ (59-3) + (114-5) + (265-7) + (306-9))/5 = 747/5 = 149.4
Average Waiting time
$\quad$ = (0+ (27-3) + (59-5) + (114+69-7) + (183+13+0-9))/5
$\quad$ = (0+24+54+176+187)/5 = 441/5 = 88.2

Table 9: Comparison between RR, DQRRR, IRRVQ, SARR, RP-5, MRR, DABRR for case-IV.

| Algorithm | Round Robin | DQRRR | IRRVQ | SARR | RP-5 | MRR | DABRR |
|---|---|---|---|---|---|---|---|
| Time Quantum | 25 | 27,68, 28,14 | 27,32,23, 27,28 | 27, 68, 28, 14 | 25,50, 100 | 27, 78, 28, 25 | 27,69,27, 14 |
| Context Switch | 15 | 7 | 10 | 7 | 8 | 7 | 7 |
| Average Waiting time | 144.4 | 107.2 | 98.2 | 88 | 104.4 | 90 | 88.2 |
| Average Turn Around Time | 205.6 | 168.4 | 159.4 | 149.2 | 165.6 | 151.2 | 149.4 |

### 4.3.2. Descending Order (Case V)

Table 10: Processes without Zero arrival time and Burst time in decreasing order.

| Processes | Arrival time | Burst Time |
|---|---|---|
| P1 | 0 | 95 |
| P2 | 2 | 75 |
| P3 | 4 | 60 |
| P4 | 8 | 43 |
| P5 | 16 | 26 |

**RR:**

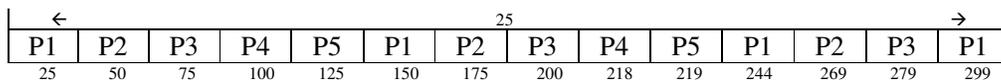

Average Turnaround time = (299+ (269-2) + (279-4) + (218-8) + (219-16))/5
$\quad\quad\quad\quad\quad$ = 1254/5 = 250.8

Average Waiting time

= ((0+100+69+35) + (25+100+69-2) + (50+100+69-4) + (75+100-8) + (100+93-16))/5
= (204+192+215+167+177)/5 = 955/5 = 191

## DQRRR:

| ←95→ | ← | 51 | → | ← 16 → | ←8→ |
|---|---|---|---|---|---|
| P1 | P5 | P2 | P4 | P3 | P2 | P3 | P2 |

95    121   172   215   266   282   291   299

Average Turnaround time = (95+ (299-2) + (291-4) + (215-8) + (121-16))/5 = 991/5 = 198.2
Average Waiting time
    = (0 + (121+94+9-2) + (215+16-4) + (172-8) + (95-16))/5
    = (0+222+227+164+79)/5 = 692/5 =138.4

## IRRVQ:

| ←95→ | ← | 26 | → | ← | 17 | → | ← 17 → | ←15→ |
|---|---|---|---|---|---|---|---|---|
| P1 | P5 | P4 | P3 | P2 | P4 | P3 | P2 | P3 | P2 |

95    121   147   173   199   216   233   250   267   284   299

Average Turnaround time = (95+ (299-2) + (267-4) + (216-8) + (121-16))/5 = 968/5 = 193.6
Average Waiting time
    = (0 + (173+34+17+0-2) + (147+43+17-4) + (121+52-8) + (95-16))/5
    = (0+222+203+165+79)/5 = 669/5 =133.8

## SARR:

| ←95→ | ← | 51 | → | ← 16 → | ←8→ |
|---|---|---|---|---|---|
| P1 | P2 | P3 | P4 | P5 | P2 | P3 | P2 |

95    146   197   240   266   282   291   299

Average Turnaround time = (95+ (299-2) + (291-4) + (240-8) + (266-16))/5 = 1161/5 = 232.2
Average Waiting time
    = (0 + (95+120+9-2) + (146+85-4) + (197-8) + (240-16))/5
    = (0+222+227+189+224)/5 = 862/5 =172.4

## R.P-5:

| ←25→ | ← | 50 | → | ← | 100 | → |
|---|---|---|---|---|---|---|
| P1 | P1 | P2 | P3 | P4 | P5 | P1 | P2 | P3 |

25    75    125   175   218   244   264   289   299

Average Turnaround time = (264+ (289-2) + (299-4) + (218-8) + (244-16))/5 = 1284/5 = 256.8
Average Waiting time
    = ((0+169) + (75+139-2) + (125+114-4) + (175-8) + (218-16))/5
    = (169+212+235+167+202)/5 = 985/5 = 197

## MRR:

| ←95→ | ← | 49 | → | ← 25 → | ←25→ |
|---|---|---|---|---|---|
| P1 | P5 | P4 | P3 | P2 | P3 | P2 | P2 |

95    121   164   213   262   273   298   299

Average Turnaround time = (95+ (299-2) + (273-4) + (164-8) + (121-16))/5 = 922/5 = 184.4
Average Waiting time
    = (0 + (213+11+0-2) + (164+49-4) + (121-8) + (95-16))/5

= (0+222+209+113+79)/5 = 623/5 = 124.6

**DABRR:**

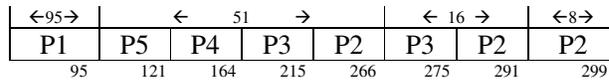

Average Turnaround time = (95+ (299-2) + (275-4) + (164-8) + (121-16))/5 = 924/5 = 184.8
Average Waiting time
= (0+ (215+9+0-2) + (164+51-4) + (121-8) + (95-16))/5
= (0+222+211+113+79)/5 = 625/5 = 125
Table 11: Comparison between RR, DQRRR, IRRVQ, SARR, RP-5, MRR, DABRR for case-V.

| Algorithm | Round Robin | DQRRR | IRRVQ | SARR | RP-5 | MRR | DABRR |
|---|---|---|---|---|---|---|---|
| Time Quantum | 25 | 95,51, 16,8 | 95,26,17, 17,15 | 95,51, 16,8 | 25,50, 100 | 95, 49, 25, 25 | 95,51, 16,8 |
| Context Switch | 13 | 7 | 10 | 7 | 8 | 7 | 7 |
| Average Waiting time | 191 | 138.4 | 133.8 | 172.4 | 197 | 124.6 | 125 |
| Average Turn Around Time | 250.8 | 198.2 | 193.6 | 232.2 | 256.8 | 184.4 | 184.8 |

### 4.3.3. Random Order (Case VI)

Table 12: Processes without Zero arrival time and Burst time in random order.

| Processes | Arrival time | Burst Time |
|---|---|---|
| P1 | 0 | 45 |
| P2 | 5 | 90 |
| P3 | 8 | 70 |
| P4 | 15 | 38 |
| P5 | 20 | 55 |

**RR:**

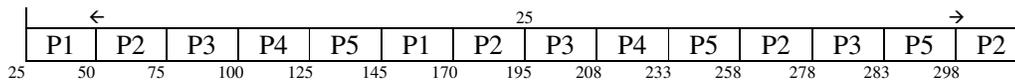

Average Turnaround time = (145+ (298-5) + (278-8) + (208-15) + (283-20))/5
= 1164/5 = 232.8

Average Waiting time
= ((0+100) + (25+95+63-25-5) + (50+95+63-8) + (75+95-15) + (100+83+45-20))/5
= (100+203+200+155+208)/5 = 866/5 = 173.2

**DQRRR:**

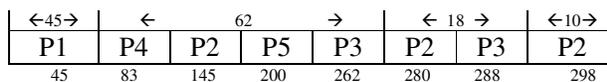

Average Turnaround time = (45+ (298-5) + (288-8) + (83-15) + (200-20))/5
= 866/5 = 173.2

Average Waiting time = (0 + (83+117+8-5) + (200+18-8) + (45-15) + (145-20))/5
    = (0+203+210+30+125)/5 = 568/5 = 113.6

**IRRVQ:**

| ←45→ | ← | 38 | → | ← | 17 | → | ← | 15 | → | ←20→ |
|------|------|------|------|------|------|------|------|------|------|------|
| P1 | P4 | P5 | P3 | P2 | P5 | P3 | P2 | P3 | P2 | P2 |
| 45 | 83 | 121 | 159 | 197 | 214 | 231 | 248 | 263 | 278 | 298 |

Average Turnaround time = (45+ (298-5) + (263-8) + (83-15) + (214-20))/5
        = 855/5 = 171
Average Waiting time = (0 + (159+34+15+0-5) + (121+55+17-8) + (45-15) + (83+76-20))/5
    = (0+203+185+30+139)/5 = 557/5 = 111.4

**SARR:**

| ←45→ | ← | 54 | → | ← | 16 | → | ←20→ |
|------|------|------|------|------|------|------|------|
| P1 | P2 | P3 | P4 | P5 | P2 | P3 | P5 | P2 |
| 45 | 99 | 153 | 191 | 245 | 261 | 277 | 278 | 298 |

Average Turnaround time = (45+ (298-5) + (277-8) + (191-15) + (278-20))/5
        = 1041/5 = 208.2
Average Waiting time = (0 + (45+146+17-5) + (99+108-8) + (153-15) + (191+32-20))/5
    = (0+203+199+138+203)/5 = 743/5 = 148.6

**R.P-5:**

| ←25→ | ← | 50 | → | ← | 100→ |
|------|------|------|------|------|------|
| P1 | P1 | P2 | P3 | P4 | P5 | P2 | P3 | P5 |
| 25 | 45 | 95 | 145 | 183 | 233 | 273 | 293 | 298 |

Average Turnaround time = (45+ (273-5) + (293-8) + (183-15) + (298-20))/5 = 1044/5 = 208.8
Average Waiting time
    = ((0+0) + (45+138-5) + (95+128-8) + (145-15) + (183+60-20))/5
    = (0+178+215+130+223)/5 = 746/5 = 149.2

**MRR:**

| ←45→ | ← | 52→ | ← | 35→ | ←25→ |
|------|------|------|------|------|------|
| P1 | P4 | P5 | P3 | P2 | P5 | P3 | P2 | P2 |
| 45 | 83 | 135 | 187 | 239 | 242 | 260 | 295 | 298 |

Average Turnaround time = (45+ (298-5) + (260-8) + (83-15) + (242-20))/5 = 880/5 = 176
Average Waiting time
    = (0 + (187+21+0-5) + (135+55-8) + (45-15) + (83+104-20))/5
    = (0+203+182+30+167)/5 = 582/5 = 116.4

**DABRR:**

| ←45→ | ← | 63 | → | ← | 17 | → | ←10→ |
|------|------|------|------|------|------|------|------|
| P1 | P4 | P5 | P3 | P2 | P3 | P2 | P2 |
| 45 | 83 | 138 | 201 | 264 | 271 | 288 | 298 |

Average Turnaround time = (45+ (298-5) + (271-8) + (83-15) + (138-20))/5
        = 787/5 = 157.4

Average Waiting time
    = (0+ (201+7+0-5) + (138+63-8) + (45-15) + (83-20))/5

= (0+203+193+30+63)/5 = 489/5 = 97.8

Table 13: Comparison between RR, DQRRR, IRRVQ, SARR, RP-5, MRR, DABRR for case-VI.

| Algorithm | Round Robin | DQRRR | IRRVQ | SARR | RP-5 | MRR | DABRR |
|---|---|---|---|---|---|---|---|
| Time Quantum | 25 | 45, 62, 18, 10 | 45,38, 17, 15, 20 | 45,54, 16,20 | 25, 50, 100 | 45, 52, 35, 25 | 45, 63, 17, 10 |
| Context Switch | 13 | 7 | 10 | 8 | 8 | 8 | 7 |
| Average Waiting time | 173.2 | 113.6 | 111.4 | 148.6 | 149.2 | 116.4 | 97.8 |
| Average Turn Around Time | 232.8 | 173.2 | 171 | 208.2 | 208.8 | 176 | 157.4 |

## 4.3.4. Comparison

This section provides the comparative analysis of seven algorithms on the basis of their resulted waiting time and turnaround time. Table 14 shows the performance analysis of seven algorithms by summarizing the waiting time and turnaround time resulted from case IV, V, VI.

Table 14: Comparison of six algorithms by aggregating their performance measure from case IV, V, VI.

| | Without 0 Arival Time | | | | | | | | | | | |
|---|---|---|---|---|---|---|---|---|---|---|---|---|
| | Context Switch | | | | Waiting Time | | | | Turnaround Time | | | |
| | Ascending Burst Time | Descending Burst Time | Random Burst Time | TOTAL | Ascending Burst Time | Descending Burst Time | Random Burst Time | TOTAL | Ascending Burst Time | Descending Burst Time | Random Burst Time | TOTAL |
| **RR** | 15 | 13 | 13 | **41** | 144.40 | 191.00 | 173.20 | **508.60** | 205.60 | 250.80 | 232.80 | **689.20** |
| **DQRRR** | 7 | 7 | 7 | **21** | 107.20 | 138.40 | 113.60 | **359.20** | 168.40 | 198.20 | 173.20 | **539.80** |
| **IRRVQ** | 10 | 10 | 10 | **30** | 98.20 | 133.80 | 111.40 | **343.40** | 159.40 | 193.60 | 171.00 | **524.00** |
| **SARR** | 7 | 7 | 8 | **22** | 88.00 | 172.40 | 148.60 | **409.00** | 149.20 | 232.20 | 208.20 | **589.60** |
| **RP-5** | 8 | 8 | 8 | **24** | 104.40 | 197.00 | 149.20 | **450.60** | 165.60 | 256.80 | 208.80 | **631.20** |
| **MRR** | 7 | 7 | 8 | **22** | 90.00 | 124.60 | 116.40 | **331.00** | 151.20 | 184.40 | 176.00 | **511.60** |
| **DABRR** | 7 | 7 | 7 | **21** | 88.20 | 125.00 | 97.80 | **311.00** | 149.40 | 184.80 | 157.40 | **491.60** |

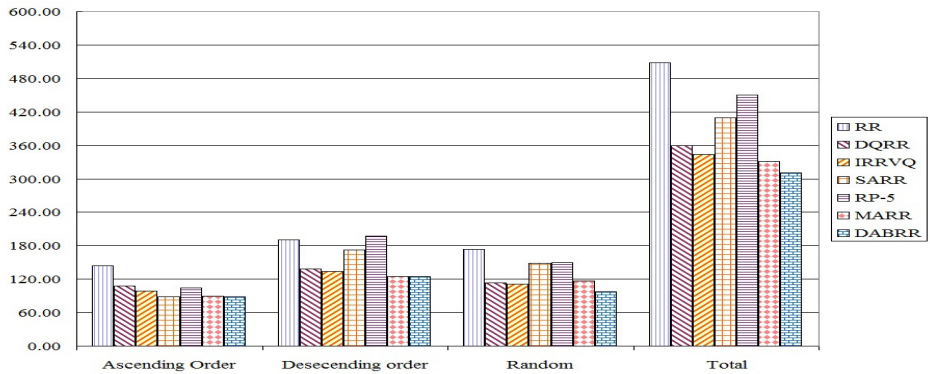

Figure 4: Waiting time for processes without 0 arrival time.

Figure 4 & 5 show the graphical analysis of the waiting time and turnaround time respectively of all algorithms based on case IV, V, VI. By evaluating these two graphs we conclude that our algorithm performs better than other six algorithms in case of processes arrived with out zero arrival time.

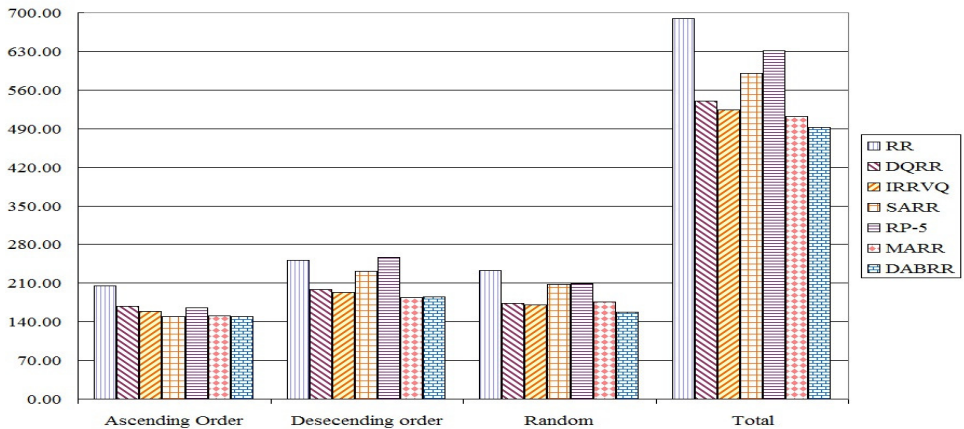

Figure 5: Turnaround time for processes without 0 arrival time.

## 5. RESULT ANALYSIS

From the analysis of all the algorithms it is concluded that our algorithm performs better than other compared algorithms in all cases.

Table 15: Percentage of waiting time reduced by each algorithm.

| Waiting Time | | | | | | | |
|---|---|---|---|---|---|---|---|
| | RR | DQRRR | IRRVQ | SARR | RP-5 | MRR | DABRR |
| Grand Total | 1,155.40 | 859.00 | 843.60 | 891.40 | 1,080.20 | 730.80 | 679.00 |
| Percentage Gain | 0.00% | 25.65% | 26.99% | 22.85% | 6.51% | 36.75% | 41.23% |

Figure 6 depicts the percentage of waiting time saved by each algorithm than traditional Round Robin algorithm. Proposed algorithm "DABRR" saves 41% waiting time as compared to traditional Round Robin.

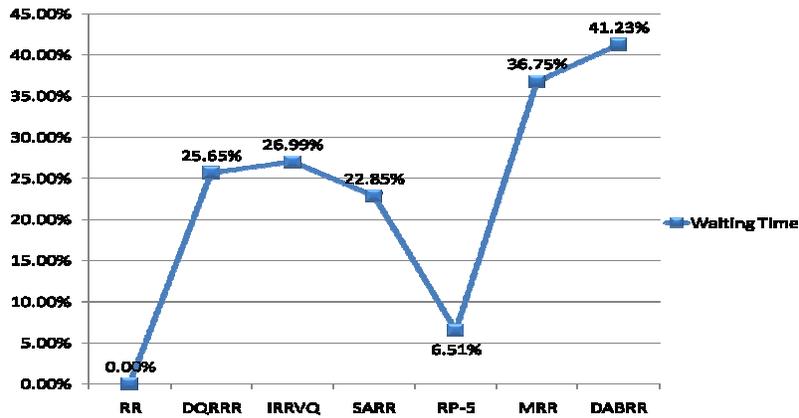

Figure 6: Percentage of waiting time reduced.

Table 16: Percentage of turnaround time reduced by each algorithm.

| Turnaround Time | | | | | | | |
|---|---|---|---|---|---|---|---|
| | RR | DQRRR | IRRVQ | SARR | RP-5 | MRR | DABRR |
| Grand Total | 1,551.60 | 1,255.20 | 1,239.80 | 1,287.60 | 1,476.40 | 1,127.00 | 1,075.20 |
| Percentage Gain | 0.00% | 19.10% | 20.10% | 20.10% | 4.85% | 27.37% | 30.70% |

Figure 7 depicts the percentage of turnaround time saved by each algorithm than traditional Round Robin algorithm. Proposed algorithm "DABRR" saves 31% turnaround time as compared to traditional waiting time.

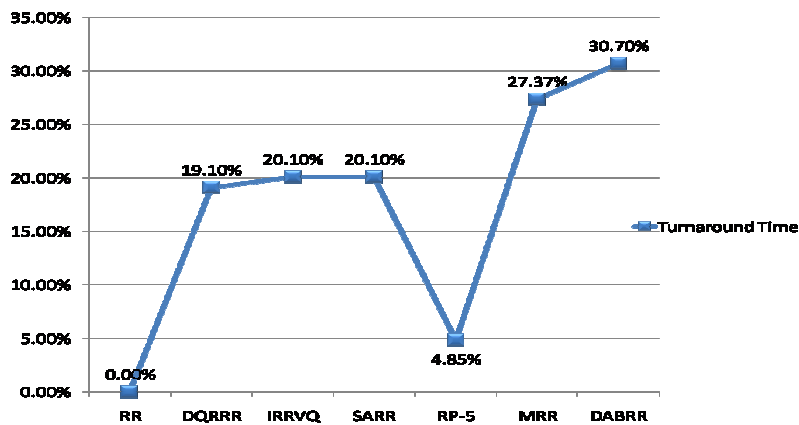

Figure 7: Percentage of turnaround time reduced.

## 6. CONCLUSIONS

This paper presents a variant of Round Robin scheduling algorithm. Comparative analysis of various algorithms like RR, DQRRR, IRRVQ, SARR, RP-5, MRR, and the proposed algorithm DABRR has been done. The proposed algorithm provides better performance metrics than the above discussed algorithms by minimizing the average waiting time and average turnaround time. In future we want to improve this algorithm for multiprocessor environment.

**Authors**

**Amar Ranjan Dash** has achieved his B.Tech. degree from Biju Patnaik University of Technology, Odisha, India and M. Tech. degree from Berhampur University, Odisha, India. His research interests include CPU Scheduling, Web Accessibility, and Cloud Computing.

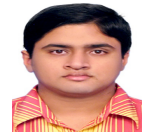

**Sandipta Kumar Sahu** has achieved his B.Tech. degree from Biju Patnaik University of Technology, Odisha, India and currently pursuing his M. Tech. degree in computer science and engineering at National Institute of Science And Technology, Odisha, India. His research interests include Operating System, Software engineering, and Computer Architecture.

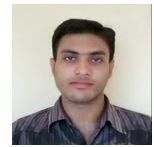

**Sanjay Kumar Samantra** has achieved his MCA degree from Berhampur University, Odisha, India and currently pursuing his M. Tech. degree in computer science and engineering at National Institute of Science And Technology, Odisha, India. His research interests include CPU scheduling, Grid computing, and Cloud Computing.

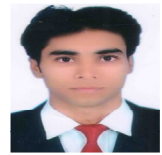